\documentclass[12pt]{article}
\usepackage{epsfig}
\usepackage{axodraw}
\usepackage{amsmath}
\def\Journal#1#2#3#4{{#1}{\bf #2} (#3) #4}

\def\PLB{{\em Phys. Lett.~}{\bf  B}}
\def\PL{{\em Phys. Lett.~}}
\def\PRL{\em Phys. Rev. Lett.~}
\def\PRD{{\em Phys. Rev.~}{\bf D}}
\def\ZPC{{\em Z. Phys.~}{\bf C}}
\def\ZP{\em Z. Phys.~}

\def\PRP{\em Phys. Rep.~}

\def\EPJC{{\em Eur. Phys. J.~}{\bf C}}

\newcommand{\X}{\mathbf{X}}
\newcommand{\e}{\mathrm{e}}
\newcommand{\f}{\mathrm{f}}

\newcommand{\p}{\mathrm{p}}
\newcommand{\q}{\mathrm{q}}
\newcommand{\fbar}{\mathrm{\overline{f}}}

\newcommand{\qbar}{\mathrm{\overline{q}}}
\newcommand{\ccbar}{\mathrm{c\overline{c}}}

\newcommand{\gast}{\gamma^*}
\newcommand{\ga}{\gamma}
\newcommand{\ra}{\rightarrow}
\newcommand{\lele}{lepton-lepton}
\newcommand{\xBj}{x_\mathrm{Bj}}

{\end{list}}
\newcounter{enumct}

\newlength{\abstwidth}
\newlength{\captivewidth}
\newcommand{\captive}[1]{\rule{5mm}{0mm}%
\begin{minipage}{\captivewidth}%
\caption[small]{#1}\end{minipage}}
 
\begin{document}
 
\sloppy
 
\pagestyle{empty}
 
\begin{flushright}
LU TP 99--35 \\
hep-ph/9911444\\
November 1999
\end{flushright}
 
\vspace{\fill}
 
\begin{center}
\boldmath
{\LARGE\bf $\gamma\gamma$ Physics at Linear Colliders}\\[10mm]
\unboldmath
{\Large Christer Friberg\footnote{christer@thep.lu.se }}\\ [2mm]
{\it Department of Theoretical Physics,}\\[1mm]
{\it Lund University, Lund, Sweden}
\end{center}
 
\vspace{\fill}
\begin{center}
{\bf Abstract}\\[2ex]
\begin{minipage}{\abstwidth}
A high-energy \lele\   collider will give us a unique possibility
to study $\e\ga$ and $\ga\ga$ interactions at high energies. The high-energy
photons can be generated by Compton back-scattering of laser light on the
high-energy lepton beams. 
With slightly reduced luminosities for $\e\ga$ and $\ga\ga$ collisions 
compared to the \lele\  collider one, unique and complementary
studies can be performed. A linear \lele\  collider also offers other photon 
sources that will be considered. 
We will focus on the complex properties of the photon itself and discuss some 
possibilities to gain new insight in its interactions at high energies and 
how to reveal the structure of it.
\end{minipage}
\end{center}

\vspace{\fill}
 
\clearpage
\pagestyle{plain}
\setcounter{page}{1}

\section{Introduction}

A linear lepton-lepton collider is the natural continuation of the program 
at CERN and the Photon Physics aspects of the $\e\p$ program at DESY. 
If there will be a possibility to study high energy collisions dedicated 
(partly) to photon interactions within a near future, this may well be the 
only opportunity. 

Assuming a \lele\  collider being built, the annihilation processes will be 
overwhelmed by a background from photon processes, which clearly have to be 
taken seriously although they may not be of primary interest. A better 
understanding of the nature of the photon is thus wanted not only from a 
photon physics point of view.

With Compton laser back-scattering of photons~\cite{CLBS}, luminosities for 
$\e\ga$ and $\ga\ga$ collisions would be not much smaller than the 
lepton-lepton one at typical center-of-mass energies. 
Several processes of great interest have a much larger 
cross section when involving photons than just leptons. A $\ga\ga$ ($\e\ga$) 
collider is then not only complementary but also highly competitive to a 
\lele\  one.

$\e\ga$ and $\ga\ga$ interactions provide one of the most complex test 
grounds for QCD at high energies. Going to higher energies than now reachable
by LEP will clearly also take us to previously unexplored regions in $x$ and 
$Q^2$ of the photon structure function.
However, an overlap with LEP measurements is highly desirable and special
effort should be taken in order to obtain that. 

Some important aspects of Higgs physics can be studied in $\ga\ga$ 
interactions and with a much higher accuracy than for the \lele\  
case~\cite{LBref}. With Compton back-scattered photons it is possible to 
perform energy scanning and to focus on particularly interesting energies. 
The resonance production of a Higgs boson, Fig.~\ref{fig:ggH}, is built up 
by a loop of charged particles that couples to the Higgs.
When at right energy, the cross section for Higgs production
is much larger in the case of $\ga\ga$ than for $\e\e$. 
The decay width of the Higgs into two photons reflects the spectrum of 
heavy charged particles where the probed
masses may possibly be far above the Higgs mass itself. 
\begin{figure}
\begin{center}
\begin{picture}(110,70)(0,0)
\Photon(0,70)(30,50){5}{6} \Text(25,70)[]{\Large $\ga$}
\Line(30,50)(30,20)
\Line(30,20)(60,35)
\Line(60,35)(30,50) \Text(70,50)[]{\Large t, W, ...}
\DashLine(60,35)(100,35){4} \Text(80,25)[]{\Large H}
\Photon(0,0)(30,20){5}{6} \Text(25,-5)[]{\Large $\ga$}
\end{picture}
\end{center}
\captive{Resonance production of a Higgs boson.
\label{fig:ggH}}
\end{figure}

In the electroweak sector, multiple gauge boson couplings can be studied. 
The pair production process $\ga\ga \ra W^+W^-$, which has a 
much larger cross section for $\ga\ga$ than the corresponding $\e\e$ one, 
can be used to determine properties as the static magnetic and 
electric multipole moments of the $W$ bosons.

In the following, we will concentrate on the photon itself: 
its properties at high energies and
the sources of photons at a linear \lele\  collider.

\section{Sources of Photons}

There are essentially three different sources of photons at a linear \lele\  
collider: bremsstrahlung, beamstrahlung and Compton laser back-scattering. 
The latter one will probably require a separate interaction region.

High-energy accelerated charged particles radiate photons, so called 
{\em bremsstrahlung} photons, which are often utilized at $\e^+\e^-$ 
colliders for doing $\ga\ga$ physics, Fig.~\ref{fig:BS}. 
It offers a spectrum of different 
photon energies and virtualities. The distributions are peaked at the lower 
end so studies of the interesting collisions of energetic and highly virtual 
photons are limited by statistics. Collisions of almost real photons can be 
isolated by antitagging conditions of the outgoing leptons. The method suffers 
from not knowing the invariant mass of the $\ga\ga$ system, which then needs 
to be reconstructed from the particles observed in the detector. 

\begin{figure}
\begin{center}
\begin{picture}(200,200)(0,0)
\ArrowLine(10,150)(80,150)
\Text(35,160)[]{{\Large $\e$}}
\ArrowLine(80,150)(160,190)
\ArrowLine(10,50)(80,50)
\Text(35,40)[]{\Large $\e$}
\ArrowLine(80,50)(160,10)
\Photon(80,150)(130,110){5}{6}
\Text(90,120)[]{\Large $\ga$}
\Photon(80,50)(130,90){5}{6}
\Text(90,80)[]{\Large $\ga$}
\GOval(135,100)(20,10)(0){0.5}
\LongArrow(143,110)(180,130)
\LongArrow(144,105)(180,115)
\LongArrow(145,100)(180,100)\Text(190,100)[]{\Large $\X$}
\LongArrow(144,95)(180,85)
\LongArrow(143,90)(180,70)
\end{picture}  
\end{center}
\captive{$\ga\ga$ process induced by bremsstrahlung from the incoming 
leptons.
\label{fig:BS}}
\end{figure}

A beam of charged particles may, in the presence of an external field, 
such as another beam of charged particles, coherently emit real photons. 
This is refered to as {\em beamstrahlung} and is a draw-back for the 
normal $\e^+\e^-$ program, wherefore \lele\  colliders are designed to 
minimize this effect. Depending on specific design parameters, such as 
the beam geometry, the beamstrahlung spectrum can be made much harder 
than the corresponding bremsstrahlung one~\cite{beamstrl}.

{\em Compton laser back-scattering} provides an intense beam of high-energy 
photons. The idea is to have a laser beam incident on a 
high-energy lepton beam. With sufficient laser flux, almost all of the energy 
of the incident lepton will be transferred, via the Compton process, to the 
scattered photon. If the invariant mass of the incident lepton and the 
laser photon system is too high, the pair production process 
$\e + \ga_\mathrm{laser} \ra \e+\e^+ + \e^-$ will occur. Similarly, a too high 
invariant mass of the scattered photon and the laser photon system will
allow the process $\ga + \ga_\mathrm{laser} \ra \e^+ + \e^-$. Both processes
can be suppressed by tuning the frequency of the laser; the latter of the 
two processes is often the one setting the upper limit. 

\begin{figure}
   \begin{center}
     \mbox{\psfig{figure=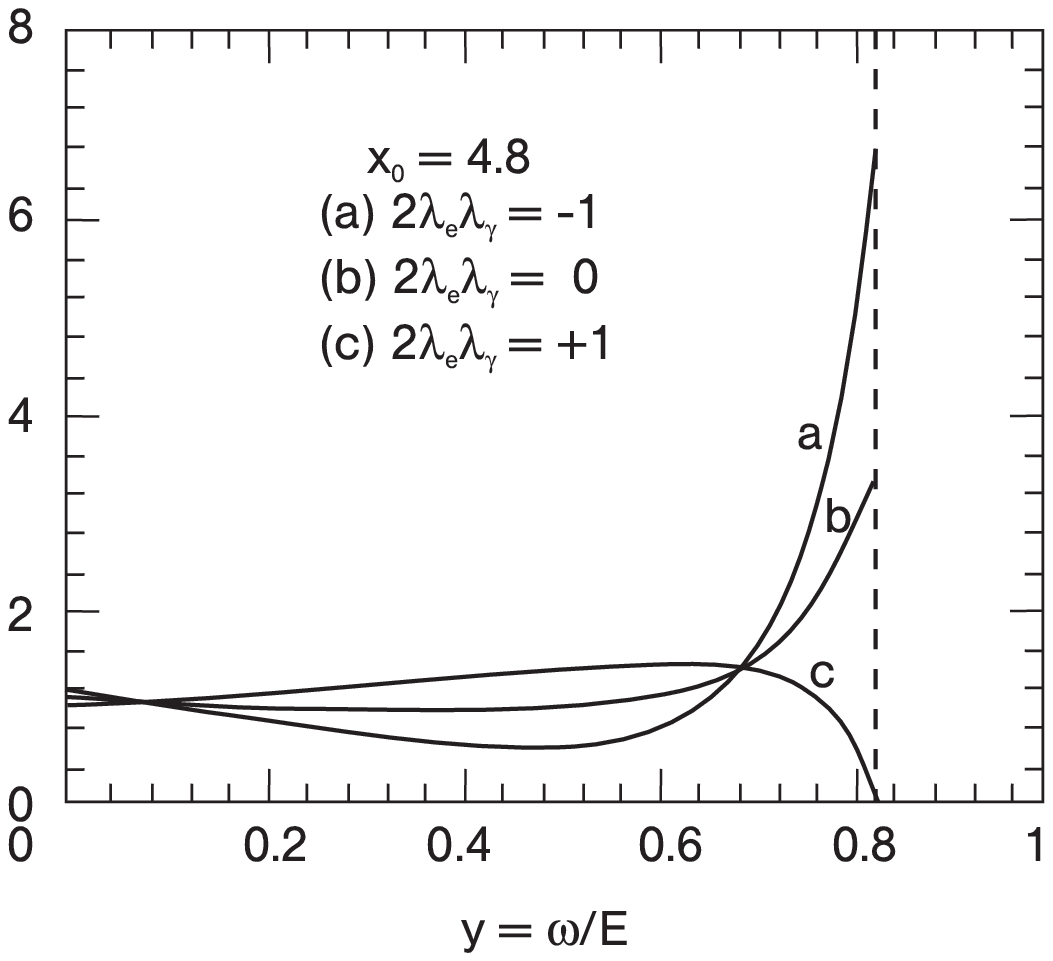,width=79mm}}
     \mbox{\psfig{figure=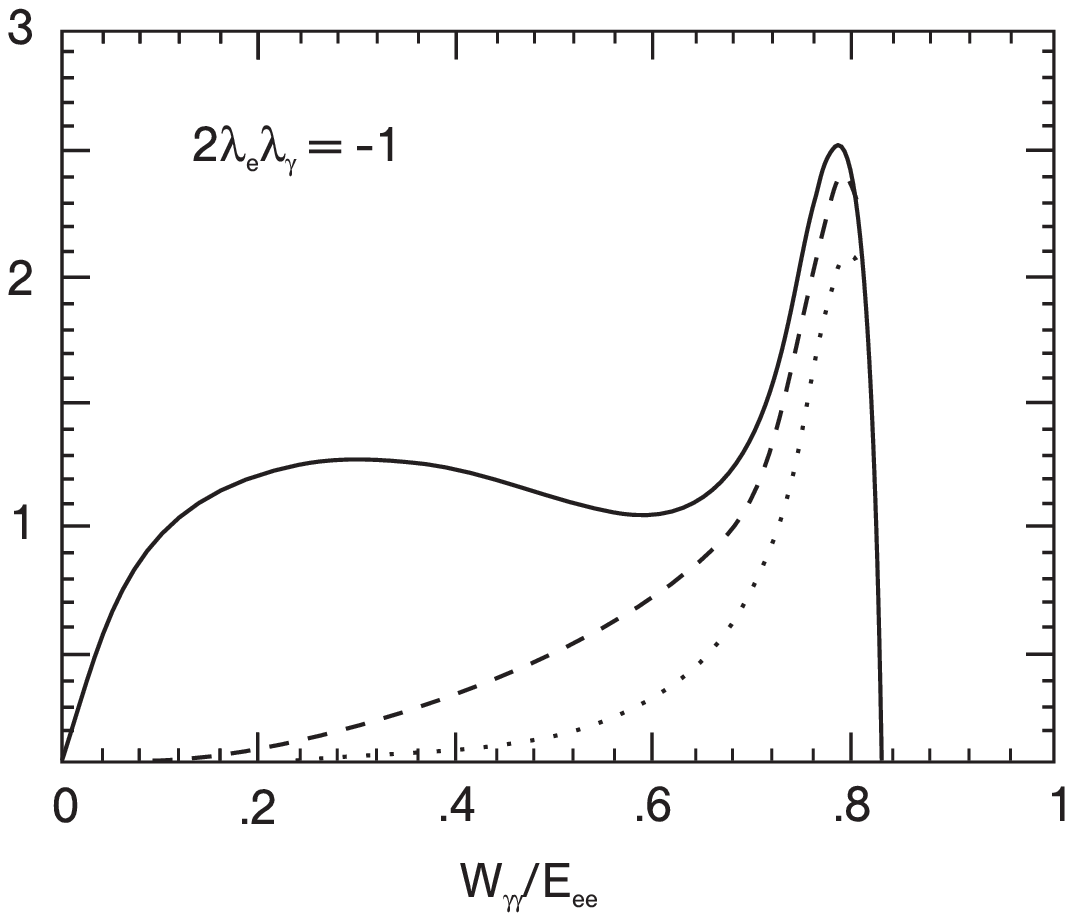,width=79mm}}
   \end{center}
\captive{To the left, the energy fraction spectrum of the Compton laser 
back-scattered photons for different helicities of the incident lepton and 
laser photon beams~\cite{LBref}. To the right, the invariant mass distribution 
of the $\ga\ga$ system for the case of opposite helicities of the lepton and 
laser beams. The dashed lines show the distribution when the distance 
between the conversion and collision point are separated.
\label{fig:LBspec}}
\end{figure}

The energy spectrum of the high-energy photons is rather broad for 
unpolarized beams. For polarized beams it can be made much sharper, as 
illustrated in Fig.~\ref{fig:LBspec}. The case of opposite helicities of 
the incident lepton and laser beams give a peaked distribution towards the 
upper end. 	

There is a strong energy--angle correlation for the Compton scattered photons.
By varying the distance between the conversion and collision point of the 
photons, a very peaked invariant mass spectrum for the $\ga\ga$ system can be 
obtained, Fig.~\ref{fig:LBspec}. 

\section{Nature of Photon}

The photon is often refered to as not having any internal structure. 
This is true for highly virtual photons, as in DIS of $\e\p$. 
For a real photon this pointlike behaviour corresponds to one of its 
components, called the direct photon. At high energies the photon may 
also fluctuate into a virtual $\q\qbar$ pair which then may undergo 
strong interactions in a collision, this is the so-called resolved photon. 

The resolved photon may be further divided into low- and high-virtuality
fluctuations. The high-virtuality ones, the anomalous photons, are calculable 
from perturbative QCD whereas the low-virtuality fluctuations are not. 
Instead, they can be approximated by Vector Meson Dominance models (VMD), 
where the photon couples directly to a vector meson state which has the 
same quantum numbers as the photon, i.e. $\rho$, $\omega$, $\phi$, etc. 
The couplings of the photon to the vector mesons are determined 
experimentally. It is then necessary with an effective description of 
the photon in terms of an {\it a priori} unknown parton 
distribution~\cite{pdf}.

In common for the resolved photon components is that they leave a beam remnant 
behind in a collision. At leading order, all of the energy from the direct 
photon goes into the hard interaction. This is reflected in the $x_\ga$ 
observable, which is defined as the fraction of the photon lightcone momentum 
that goes into the production of jets. Direct photon events have $x_\ga=1$ 
but various effects smear out the distribution to lower values. 

A hard-scattering $\ga\ga$ process can then be classified according to 
the number of photons being resolved, Fig.~\ref{fig:gg}. The direct process  
$\ga\ga \ra \f\fbar$ is the ordinary fermion pair production process.
With one or both photons resolved, 
the hard scattering-processes should be convoluted with the parton 
distribution of the resolved photon. The single-resolved processes 
correspond to the direct ones in $\ga\p$, and the double-resolved ones to
resolved processes in $\ga\p$. 

\begin{figure}
\begin{center}
\begin{picture}(382,100)(0,0)
  \Photon(10,90)(45,70){4}{3}   \ArrowLine(85,90)(45,70)
  \Photon(10,10)(45,30){4}{3}   \ArrowLine(45,30)(85,10)
      \Line(45,70)(45,30)
     \Vertex(45,70){2}   \Vertex(45,30){2}
    \Text(1,90)[r]{\Large $\ga$}
    \Text(1,10)[r]{\Large $\ga$}
     \Photon(150,90)(200,75){4}{5}   \ArrowLine(200,75)(230,90)
     \Photon(150,10)(180,20){4}{3}   \ArrowLine(230,10)(180,20)
  \ArrowLine(180,20)(200,35)         \Gluon(200,35)(230,20){4}{3}    
      \Line(200,75)(200,35)
     \Vertex(200,75){2}   \Vertex(200,35){2}
    \Text(141,90)[r]{\Large $\ga$}   \Text(141,10)[r]{\Large $\ga$}
     \Photon(290,90)(320,80){4}{3}   \ArrowLine(380,90)(320,80)
  \ArrowLine(290,10)(350,30)         \ArrowLine(350,30)(380,10)
  \ArrowLine(320,80)(350,60)         \ArrowLine(350,60)(380,80)
      \Gluon(350,60)(350,30){4}{3}
     \Vertex(350,60){2}   \Vertex(350,30){2}
\DashLine(290,5)(380,5){4}
        \GOval(285,10)(10,5)(0){0.5}
    \Text(276,90)[r]{\Large $\ga$}   \Text(276,10)[r]{\Large $\ga$}
\end{picture}
\end{center}
\captive{Examples of a direct, 
a single-resolved (a direct and an anomalous photon) 
and a double-resolved (an anomalous and a VMD photon) 
event in $\ga\ga$ collisions.
\label{fig:gg}}
\end{figure}

\section{$\ga\ga$ total cross sections}

\begin{figure} [ht]
   \begin{center}
     \mbox{\psfig{figure=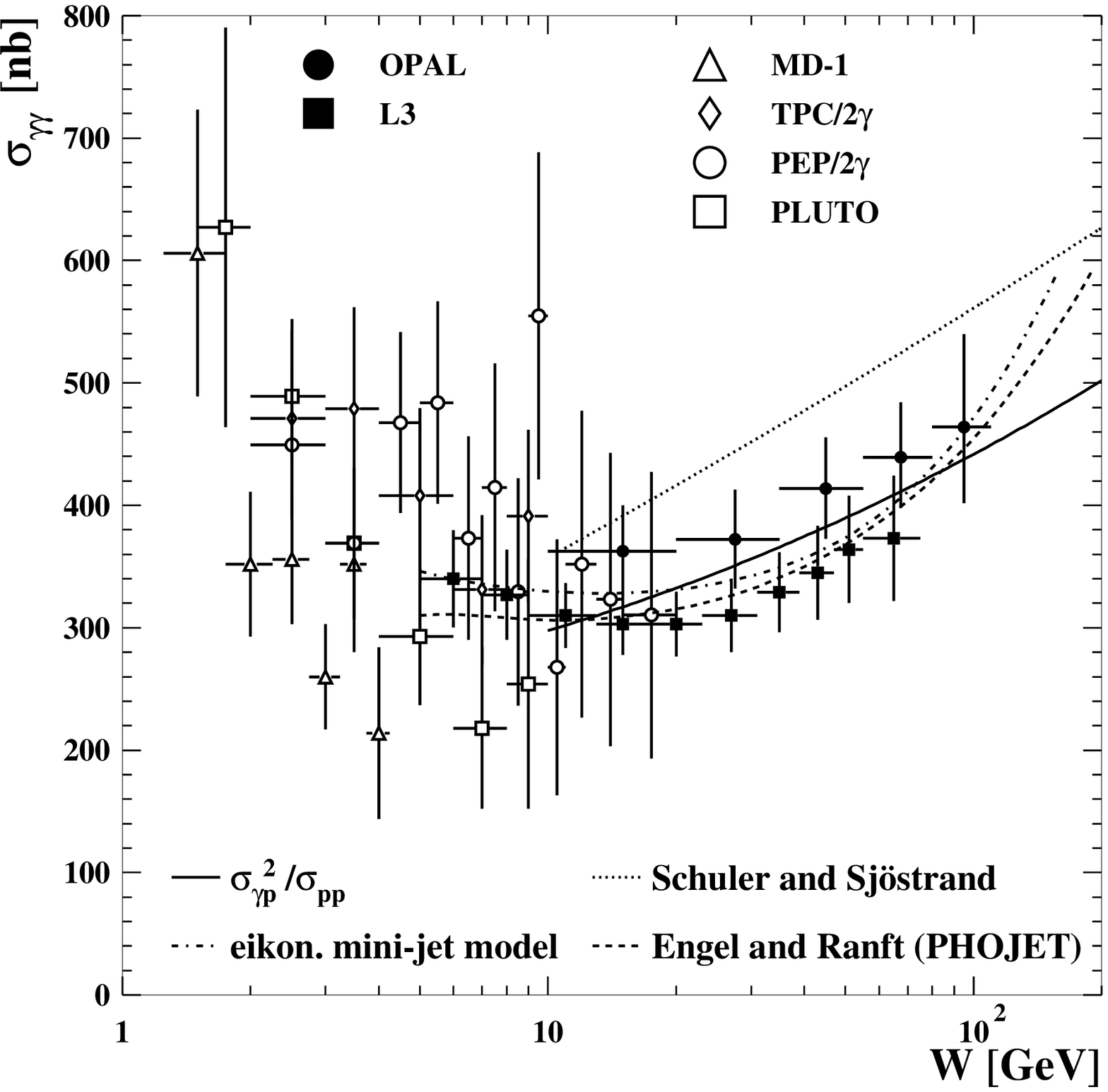,width=10cm}}
   \end{center}
\captive{$\ga\ga$ total cross section. For a description of the models, 
see~\cite{ggtot} and references therein. The model curve here labelled 
Schuler and Sj\"ostrand is intended as an upper limit, not a central value.  
\label{fig:ggtot}}
\end{figure}

The $\ga\ga$ total cross section is not known from first principles and is 
poorly measured so far~\cite{ggtot}. The situation is even worse for the 
case of mildly virtual photons. When measuring the $\ga\ga$ total cross section
from the process $\e^+\e^- \ra \e^+\e^- + \mathrm{hadrons}$, i.e. using 
bremsstrahlung photons, antitagging conditions have to be applied for the 
scattered leptons. The invariant mass $W_{\ga\ga}$ of the $\ga\ga$ system 
then has to be reconstructed from the finally observed particles. The method 
of unfolding is used for estimating the correct invariant mass from the 
measured one. 
Since a large fraction of the final-state particles go undetected, the 
estimate has to be based on a sound model that hopefully describe the correct 
total cross section. In the end it leads to large systematic errors. 
In Fig.~\ref{fig:ggtot}, OPAL have presented a result which is based on 
unfolding with two different models. The result of L3 is based on unfolding
with one model only.	

As for hadron-hadron collisions the $\ga\ga$ total cross section is rising 
with increasing energy, Fig.~\ref{fig:ggtot},  
which is consistent with a universal Regge behaviour of total cross sections. 
Since the $\ga\ga$ total cross section presumably is dominated by the 
hadronic component of the photon, important cross-checks can be done with 
$\ga$-hadron and hadron-hadron collisions. 

To obtain a better understanding of the $\ga\ga$ total cross section, it is 
important to measure it at a wide range of energies. This can be done,
in principle, by Compton back-scattered photons since they can be obtained 
within a narrow energy interval, as described in a previous section. The 
problem with unfolding is not present but the number of 
$\ga\ga$ collisions per bunch crossing may be substantial, depending on the 
actual experimental set-up, making the interpretation of data harder than
first thought of. 

Recent measurements of the $\gast\gast$ total cross section by L3 and 
OPAL is in conflict with leading order BFKL calculations~\cite{smallx,BFKL}. 
A model with the concept of virtual resolved photons describes the data 
however. It may indicate that the current $x$-values available at LEP are 
too large for the summation of $\log(x)$ terms to be relevant. 

\section{Photon Structure}

The cross section for DIS $\e\ga$ can be used to extract the $F_2^\ga$ 
structure function of the real photon. The parton distributions obey 
$Q^2$ evolution equations and can be written as a sum of two terms; 
an inhomogeneous and a homogeneous term. The (non-perturbative) 
homogeneous term is familiar from the proton parton distributions whereas 
the (perturbative) inhomogeneous one arises from the point-like coupling 
of the photon to a $\q\qbar$ pair (thereof the name, anomalous photon). 

To reveal the structure of the photon, a probing virtual photon can be 
provided by bremsstrahlung photons. Complementary studies can be obtained 
by charged currents in DIS $\e\ga$. 
Previously unexplored regions of high $Q^2$ can be reached. Moreover, with a 
small-angle tagger, an overlap with LEP measurements and previously 
unaccessible small values of Bjorken-$x$ ($\xBj$) can be obtained.
This is desirable to give constraints on the poorly known gluon 
distribution of the photon. 

The target photon can be obtained from any of the three sources previously 
discussed. They offer different kinematical regions, however. 

To study the virtual-photon structure, available from bremsstrahlung photons, 
double-tagged events is needed but these occur at low rates. Again, a 
small-angle tagger is desirable for overlap with previous measurements.

Using bremsstrahlung photons as targets will give a large systematic error
for small-$\xBj$ measurements due to the problems with an unknown invariant 
mass of the collision. At large-$\xBj$ the systematic errors are smaller and
will give complementary measurements to those at LEP (overlapping or not).

The beamstrahlung photons can be much harder than the bremsstrahlung ones, 
giving access to small-$\xBj$ values --- provided that there is a small-angle 
tagger. Furthermore, with beamstrahlung photons the event 
rates can be peaked at the lowest end of the $\xBj$ distribution giving 
reasonable statistics~\cite{beamstrl}. 

If both lepton beams are Compton back-scattered, interactions of high energy 
real photons can be studied at luminosities comparable to the \lele\  case. 
With only one beam back-scattered, deep inelastic scattering of 
$\e\ga$ can be used to measure the real photon structure. It is ideal for 
small-$\xBj$ measurements since large invariant masses are easily obtained. 
The systematic errors for large $\xBj$ will, of course, also be reduced with 
this option.

The parton content of the photon can be further explored by looking at charm 
production, $\e^+\e^- \ra \e^+\e^- + \ccbar$. At \lele\  collider energies 
of 500~GeV, the single-resolved cross section is larger than the direct one 
and since the boson-gluon fusion process dominates the single-resolved 
processes; it probes the gluon structure of the photon. 

\section{Summary}

By varying geometric parameters and the polarization of the incident lepton 
and laser beams, Compton laser back-scattering can be used to perform energy 
scanning over a broad energy range or to focus on a particularly 
interesting energies, such as the Higgs mass. The flexibilities and 
possibilities, whereof several unique ones, make a Compton collider 
competitive to the original \lele\  collider.

For measuring the photon structure, a small-angle tagger is desirable for
obtaining overlapping results with LEP. The small-$\xBj$ region is 
outstandingly best measured with the Compton back-scattering option. 
This region will give access to the gluon content of the photon and will 
offer new stringent tests of QCD. It can also be reached with beamstrahlung
photons but the performance of the measurement rely on specific design 
parameters of the collider.

The $\ga\ga$ total cross section will require a small-angle tagger to be
measured with bremsstrahlung photons. Again, this can probably be done much
better with Compton back-scattering, depending on the actual design of the 
machine. 

Although photon physics may not be of primary interest at a linear 
\lele\  collider it will be one of the major backgrounds for many 
processes of the original $\e^+\e^-$ program, and thereby an over-all better 
understanding of the photon and its interactions is wanted.

\end{document}